\documentclass[aps,pre,onecolumn,superscriptaddress,longbibliography]{revtex4-2}
\usepackage{graphicx}
\usepackage{color}
\usepackage{hyperref,breakurl,amsmath,amssymb,pst-node,eepic}
\usepackage{bm,systeme}
\DeclareMathOperator*{\argmax}{argmax}

\newcommand{\be}{\begin{equation}}
\newcommand{\ee}{\end{equation}}
\newcommand{\de}{\mathrm{d}}

\makeatletter
\let\REVTeX@orig@bibinfo\bibinfo
\def\bibinfo#1#2{%
  \def\REVTeX@tag{title}%
  \def\REVTeX@arg{#1}%
  \ifx\REVTeX@tag\REVTeX@arg
    \textit{``#2''}
  \else
    #2%
  \fi
}
\makeatother

\begin{document}

\title{A statistical mechanical view of complex economies under the green transition}

\author{Filippo Marcuccini}
\affiliation{Dipartimento di Fisica, Universit\`a di Pavia, Via Bassi 6, 27100 Pavia, Italy}

\author{Giacomo Livan}
\email{giacomo.livan@unipv.it}
\affiliation{Dipartimento di Fisica, Universit\`a di Pavia, Via Bassi 6, 27100 Pavia, Italy}
\affiliation{Istituto Nazionale di Fisica Nucleare, Sezione di Pavia, Via Bassi 6, 27100 Pavia, Italy}
\affiliation{Department of Computer Science, University College London, 66-72 Gower Street, London WC1E 6EA, UK}


\begin{abstract}
We propose a stylized model of a complex economy to explore the economic tradeoffs imposed by the so called `green transition' --- the shift towards more sustainable production paradigms --- using tools from the Statistical Mechanics of disordered systems. Namely, we promote the parameters of a standard input-output economic model to random variables, in order to characterize the typical statistical properties of its equilibria. A central feature of our work is the explicit inclusion of a waste variable as a proxy for the unwanted byproducts, such as emissions, generated by different production processes. We find that the interplay between economic development and waste gives rise to a double phase transition, separating a region of viable economic activity from two distinct shutdown regimes. Notably, more wasteful (`brown') economies support a larger number of technologies per good but fail to activate most of them, limiting their productive capacity. In contrast, `greener' economies, while having access to fewer technologies, tend to achieve higher economic welfare, up to a critical point where rapid growth under tight waste constraints may destabilize them. Our results highlight the structural tensions between technological development and environmental sustainability, suggesting that the green transition may require careful navigation to avoid abrupt collapses in economic activity. The nature of the phase transitions we observe in the model is linked to the shrinkage of feasible economic configurations under increasing constraints, akin to algorithmic phase transitions studied in Computer Science.


\end{abstract}

\maketitle

\section{Introduction} \label{sec:intro}

Modern developed economies have reached remarkable levels of complexity. Nowadays, technologically sophisticated products are delivered through intricate supply chains that often involve the interaction of hundreds, or even thousands, of heterogenoues actors, ranging from designers and developers, to factories and logistics companies~\cite{pichler2023building,carvalho2021supply}.

As is well known by now, such economic development has come at the price of significant changes to our climate~\cite{shukla2022climate}. As a consequence, our economies now face the daunting task of seeking to transition to cleaner paradigms of mass production, while at the same time retaining the economic benefits delivered by past growth. Such a challenge is often broadly referred to as the `green transition', i.e., the reorientation of modern economies towards low-emission and resource-efficient ways of delivering goods to the global population.

From a high level perspective, the green transition amounts to an unprecedented attempt at `reconfiguring' a large complex system. There is extensive debate on the pros and cons of different pathways towards greener economies~\cite{shayegh2023assessment}, due to the importance of the tradeoffs involved and the impact they may have on the lives of countless people. On the one hand, it is rather widely accepted that---on top of the obvious gains in terms of global health---the green transition will open up new economic opportunities in the medium-long term, favouring technological innovation~\cite{acemoglu2012environment} and leading to the creation of new jobs~\cite{lehr2012green}. At the same time, though, it is also understood that the transition itself will not be painless in the short run, as it will likely result in job losses and/or in significant skill mismatches in the workforce~\cite{hanna2024job}, as well as in the exacerbation of preexisting socio-economic inequalities~\cite{fremstad2019impact}. 

Getting a precise quantitative understanding of the above tradeoffs and their \emph{net} effects on the global economy is an almost hopeless task, due to practical impossibility of collecting fine-grained data about economic production networks from different countries~\cite{pichler2023building,bacilieri2023firm}, which often can only be inferred at best~\cite{mungo2023reconstructing}. Even worse, should those data be readily available, there would still be no obvious way of aggregating them into accurate predictions of the global economy's future trajectory, due to the unprecedented nature of the course we are on. 

In this paper, we seek to shed some light on the above tradeoffs by leveraging approaches that are typically used in the Statistical Mechanics of disordered systems~\cite{stein2013spin}, which provides a natural framework to characterize the \emph{average} behavior of large complex interacting systems. The key insight provided by Statistical Mechanics is that the microscopic details of a large interacting system often become irrelevant to its macroscopic statistical description, resulting in so called self-averaging behavior. Such insight has led to massive advances in our understanding of a great variety of both physical and non-physical complex systems, ranging from spin glasses~\cite{mezard1987spin} and quantum fields~\cite{verbaarschot2000random} to ecosystems~\cite{allesina2012stability}, machine learning algorithms~\cite{carleo2019machine}, and game theory~\cite{galla2013complex}, just to name a few.

In the following, we will consider a standard \emph{General Equilibrium Theory} (GET) model of an economy where profit-maximizing firms interact with each other and with utility-maximizing consumers, and seek to derive its statistical properties upon averaging over all possible interactions---within certain constraints---between such actors. Broadly speaking, GET refers to the analysis of the conditions under which supply and demand match via the interaction between rational economic agents~(see, e.g.,~\cite{lancaster2012mathematical}). Despite being increasingly criticized for their somewhat unrealistic assumptions~\cite{kirman1992whom,chang2013labor}, GET models remain very much in use among leading global economic institutions. Furthermore, a handful of studies---which our work builds upon---have already demonstrated how recasting GET models in a statistical mechanical framework leads to the emergence of results, such as sharp transitions between different economic regimes, that are in qualitative agreement with empirical observations but cannot be produced via classical GET modelling~\cite{de2005typical,de2006statistical,de2007typical,bardoscia2012financial,marsili2014complexity,bardoscia2017statistical,moran2019may}.

Within this stream of literature, our main contribution will be that of explicitly modelling and accounting for the \emph{waste} generated by an economy, which in our simplified setting represents a  proxy for, e.g., greenhouse gas emissions and other pollutants generated by modern economies. 

In Section~\ref{sec:random_economies}, we illustrate our model, and in Section~\ref{sec:sol} we present its analytical solution (which we detail in full in the Appendices). In Section~\ref{sec:results}, we present our main results, highlighting two phase transitions that the economy described by our model undergoes in certain settings. The nature of such phase transitions is detailed in Section~\ref{sec:pt}.

\section{Large Random Economies} \label{sec:random_economies}

We consider an economy with $N$ firms, $C$ goods and one representative consumer, a concept often used in GET modelling to capture average/typical consumer behaviour~\cite{lancaster2012mathematical}. Each firm is endowed with a technology that allows it to produce a certain bundle of goods by consuming others. This is captured by a matrix with entries $q_n^c$ that are positive (negative) when firm $n$ produces (consumes) good $c$ ($n = 1, \ldots, N$; $c = 1, \ldots C$). Such a matrix is somewhat akin to the tables compiled in input-output (IO) analysis~\cite{leontief1986input,miller2009input}---the branch of Economics that seeks to model the interdependencies between different sectors of an economy and/or different economies. In that analogy, $q_n^c$ would quantify, e.g., the amount of goods produced by sector (country) $n$ that sector (country) $c$ consumes/imports in order to achieve its own production goals.

The matrix entries $q_n^c$ in the above framework effectively represent \emph{technical coefficients}, expressing the amounts of goods produced and consumed by firms in some suitable unit measure. As such, one can choose a scale such that a firm's total output equals one, i.e.,
\begin{equation*}
\sum_{c=1}^C \Theta (q_n^c) = 1 \ ,
\end{equation*}
where $\Theta(x) = 1$ for $x > 0$ (and $\Theta(x) = 0$ otherwise) only selects those goods that are produced by firm $n$. With this position, we then assume that each firm generates a certain amount of waste $\epsilon_n > 0$ per output unit, i.e.,
\begin{equation} \label{eq:eps}
\sum_{c=1}^{C}q_n^c=-\epsilon_n \ .
\end{equation}
The above quantity captures the a firm's efficiency as the discrepancy between one unit of the output produced by its technology, and the input required to produce it. As such, in the following we will relate it to the undesired byproducts---most notably emissions---associated with production of technologically advanced products on a large scale.

Firms operate their technologies at a scale $s_n > 0$, with $s_n = 0$ indicating that firm $n$ is inactive. When active, firms make a profit
\begin{equation*}
\pi_n = s_n\sum_{c=1}^{C}q_n^c p^c = s_n \ \boldsymbol{p} \cdot \boldsymbol{q}_n \ ,
\end{equation*}
where $\boldsymbol{p} = (p^1, \ldots, p^C)$ is the set of equilibrium prices for all goods ($p^c \geq 0, \ \forall \ c$)  and $\boldsymbol{q}_n = (q_n^1, \ldots, q_n^C)$.

The representative consumer's welfare is described by a utility function 
\begin{equation} \label{eq:utility}
U_N(\boldsymbol{x}) = \sum_{c=1}^C u(x^c) \ ,
\end{equation}
where 
\begin{equation} \label{eq:xc}
x^c = x_0^c + \sum_{n=1}^N s_n q_n^c \geq 0
\end{equation}
denotes the amount of good $c$ consumed at equilibrium. In the above expression, $x_0^c$ represent the economy's initial endowment of good $c$, i.e., goods for which $x_0^c > 0$ are primary goods (e.g., crops, raw materials), whereas goods for which $x_0^c = 0$ are entirely produced via technologies. The term $\sum_{i=1}^N s_n q_n^c$ quantifies the net production/consumption of good $c$ by the firms' technologies. In the following, we will denote the fraction of primary goods in the economy as $f_0$, i.e., 
\begin{equation} \label{eq:f0}
f_0 = \frac{1}{C} \sum_{c=1}^C \Theta (x_0^c) \ ,
\end{equation}
where $\Theta(x) = 1$ for $x > 0$, and $\Theta(x) = 0$ otherwise.

The economy's equilibrium prices $\boldsymbol{p}$ are determined by jointly solving two optimization problems. On the supply side, firms seek to maximize their profits by determining optimal scales of production, i.e, each firm seeks to solve $\max_{s_n \ge 0}\pi_n$. On the demand side, the representative consumer seeks to maximize utility compatibly with a budget constraint: $\max_{\boldsymbol{x} \in B} U_N(\boldsymbol{x})$, where $B = \{ x^c \geq 0 \ : \ \boldsymbol{p} \cdot \boldsymbol{x} \leq \boldsymbol{p} \cdot \boldsymbol{x}_0 \}$ denotes the set of feasible consumption bundles, i.e., non-negative consumption vectors $\boldsymbol{x} = (x^1, \ldots, x^C)$ (see Eq.~\eqref{eq:xc}) that do not require the consumer to spend more than they would without any active technology.

It can be shown~\cite{lancaster2012mathematical,de2007typical,marsili2014complexity,bardoscia2017statistical} that the joint solution of the two above optimization problems, coupled with the condition in Eq.~\eqref{eq:xc}, is equivalent to solving
\begin{equation} \label{eq:opt_scales}
\boldsymbol{s}^\star = \argmax_{\boldsymbol{s} \geq 0} \ U_N(\boldsymbol{s} | \boldsymbol{q}, \boldsymbol{\epsilon}, \boldsymbol{x}_0) = \argmax_{\boldsymbol{s} \geq 0} \ U_N \left ( \boldsymbol{x}_0 + \sum_{n=1}^N s_n \boldsymbol{q}_n \right ) \ ,
\end{equation}
which determines the optimal scale of production of each technology $\boldsymbol{s}^\star = (s_1^\star, \ldots, s_N^\star)$ and, via Eq.~\eqref{eq:xc}, the optimal consumption bundle for the consumer $\boldsymbol{x}^\star = (x_1^\star, \ldots, x_C^\star)$ for a \emph{fixed} production matrix $\boldsymbol{q} = \{ q_n^c \}_{n= 1,\ldots,N}^{c = 1,\ldots, C}$, waste vector $\boldsymbol{\epsilon} = (\epsilon_1, \ldots, \epsilon_N)$ (see Eq.~\eqref{eq:eps}), and initial endowment $\boldsymbol{x}_0$.

Our goal in the following is to characterize the economy's properties irrespective of its details. In order to do so, we will leverage ideas and methods borrowed from the Statistical Mechanics of disordered systems. Namely, we will `promote' $\boldsymbol{q}$, $\boldsymbol{\epsilon}$ and $\boldsymbol{x}_0$ to random variables, and seek to describe the economy's statistical properties when it grows large, i.e., when both the number of firms and goods grow to infinity at a fixed rate, $N, C \rightarrow \infty$ with $\nu = N/C$ fixed. In Statistical Mechanics, this is known as the thermodynamic limit. Under such limit---as is often the case with large interacting systems---the economy's main statistical properties will display self-averaging behaviour and attain typical values that do not depend on the economy's specificities. In a nutshell, this amounts to 
\begin{equation} \label{eq:htheta}
\lim_{N \rightarrow \infty} \frac{1}{N} \max_{\boldsymbol{s} \geq 0} \ U_N(\boldsymbol{s} | \boldsymbol{q}, \boldsymbol{\epsilon}, \boldsymbol{x}_0) =
\lim_{N \rightarrow \infty} \frac{1}{N} \left \langle \max_{\boldsymbol{s} \geq 0} \ U_N(\boldsymbol{s} | \boldsymbol{q}, \boldsymbol{\epsilon}, \boldsymbol{x}_0) \right \rangle_{\boldsymbol{q}, \boldsymbol{\epsilon}, \boldsymbol{x}_0} \ ,
\end{equation}
where $\langle \cdots \rangle_{\boldsymbol{q}, \boldsymbol{\epsilon}, \boldsymbol{x}_0}$ denotes averaging over the probability density functions describing the statistical properties of the economy's production matrix, waste, and initial endowment (which we will specify in the next Section). The above result, coupled with Eq.~\eqref{eq:opt_scales}, ensures that all economies will `look alike', as long as they are large enough.

In order to keep the paper's focus on our main results, we briefly outline the procedure to solve the above optimization problem in Appendix~\ref{app:statmech}, which the reader well versed in the Statistical Mechanics of disordered systems can safely skip. The reader interested in a thorough and considerably more exhaustive illustration than the sketch provided in Appendix~\ref{app:statmech} is referred to the excellent review in~\cite{castellani2005spin}.

\section{Solution and typical properties} \label{sec:sol}

In Appendix~\ref{app:sp} we provide a detailed analytical solution to the optimization problem in Eq.~\eqref{eq:htheta} under the following assumptions:
\begin{itemize}
\item As customary~\cite{lancaster2012mathematical}, we choose the utility function in Eq.~\eqref{eq:utility} to be concave and monotonically increasing by choosing $u(x^c) = \log(x^c)$
\item We assume the amount of waste generated by each technology (see Eq.~\eqref{eq:eps}) to be distributed according to a Gaussian distribution with mean $\epsilon_0$ and variance $\lambda > 0$:
\begin{equation} \label{eq:waste_pdf}
P(\epsilon_n) = \frac{1}{\sqrt{2\pi\lambda}} \mathrm{e}^{-\frac{(\epsilon_n - \epsilon_0)^2}{2\lambda}} \ .
\end{equation}
Throughout the rest of the paper we choose $\epsilon_0 = g \lambda$ with $g = 10$ in order to effectively make sure that $\epsilon_n > 0$, as it should be. Albeit inelegant, this assumption is a necessary compromise to retain the Gaussian form of the above probability distribution, as it is often prohibitive to make any analytical progress in replica trick calculations---the analytical methodology we are using in order to average over economies, see Appendices~\ref{app:statmech} and~\ref{app:sp}---with non-Gaussian distributions. Moreover, let us also emphasise that one can safely expect the final results not to change qualitatively as long as the pdf in Eq.~\eqref{eq:waste_pdf} remains defined on an effectively positive support with variance proportional to $\lambda$. In this respect, the value of $g$ does not play any relevant role, as long as it ensures that such conditions are met.

\item We assume each row of the production matrix to be drawn from a Gaussian distribution with mean zero and variance $1/C$. This, however, would not guarantee that the constraint in Eq.~\eqref{eq:eps} is met for fixed $\epsilon_n$ (drawn from the pdf detailed in the above point). In order to make sure of that, we enforce the constraint on each technology with a Delta function, resulting in the following pdf:
\begin{equation} \label{eq:IO_pdf}
P(\boldsymbol{q}_n) = \frac{\prod_{n=1}^N \delta \left ( \sum_{c=1}^C q_n^c + \epsilon_n \right ) \mathrm{e}^{-\frac{C}{2}(q_n^c)^2}}
{\int_{-\infty}^\infty \de \boldsymbol{q}_n^\prime \ \prod_{n=1}^N \delta \left ( \sum_{c=1}^C (q_n^c)^\prime + \epsilon_n \right ) \mathrm{e}^{-\frac{C}{2}(q_n^c)^{\prime 2}}} \ ,
\end{equation}
where the denominator ensures proper normalization. All in all, the above form guarantees that the vectors $\boldsymbol{q}_n$ are normally distributed multivariate random variables that exactly match the desired constraints.

\item We assume each good to be primary with probability $f_0$ (see Eq.~\eqref{eq:f0}), and, if so, to have unit initial endowment:
\begin{equation}  \label{eq:f0_pdf}
P(x_0^c) = f_0 \delta(x_0^c - 1) + (1-f_0) \delta(x_0^c) \ .
\end{equation}
\end{itemize}

With these positions, the solution to Eq.~\eqref{eq:htheta} reads
\begin{equation*}
\lim_{N \rightarrow \infty} \frac{1}{N} \left \langle \max_{\boldsymbol{s} \geq 0} \ U_N(\boldsymbol{s} | \boldsymbol{q}, \boldsymbol{\epsilon}, \boldsymbol{x}_0) \right \rangle_{\boldsymbol{q}, \boldsymbol{\epsilon}, \boldsymbol{x}_0} = 
\lim_{N \rightarrow \infty} \frac{1}{N} \left \langle \max_{\boldsymbol{s} \geq 0} \ U_N \left (\boldsymbol{x}_0 + \sum_{n=^N} s_n \boldsymbol{q}_n \right ) \right \rangle_{\boldsymbol{q}, \boldsymbol{\epsilon}, \boldsymbol{x}_0} = h(\Omega^\star,\kappa^\star,p^\star,\sigma^\star,\chi^\star,\hat{\chi}^\star) \ ,
\end{equation*}
where $h$ is a function which we compute in Appendix~\ref{app:sp}. Its arguments are the so called \emph{order parameters}, which can be identified with the statistical mechanical approach we outline in Appendix~\ref{app:statmech}, and explicitly evaluated as the solution to a set of coupled non-linear equations, known as \emph{saddle point equations} (see Eq.~\eqref{eq:sp} in Appendix~\ref{app:sp}).

\section{Results} \label{sec:results}

In Appendix~\ref{app:sp} we detail the explicit form of the saddle point equations in Eq.~\eqref{eq:sp}, which must be solved numerically. Their solution provides access to a number of key quantities (see Appendix~\ref{app:stat_prop}) that capture the typical behaviour of the economy as a function of its main parameters. More specifically, in the following we present our results focusing on ($i$) the ratio $\nu = N/C$ (number of available technologies per good), which we take as a proxy for the economy's development level; ($ii$) the $\lambda$ parameter that characterizes both the average and the variation in the distribution of waste generated by different technologies (in short, the higher $\lambda$ the more the economy's technologies generate waste, see Eq.~\eqref{eq:waste_pdf}); ($iii$) the fraction $f_0$ of primary goods in the economy (Eq.~\eqref{eq:f0}).

\begin{figure}[h!]
\centering
\includegraphics[width=\textwidth]{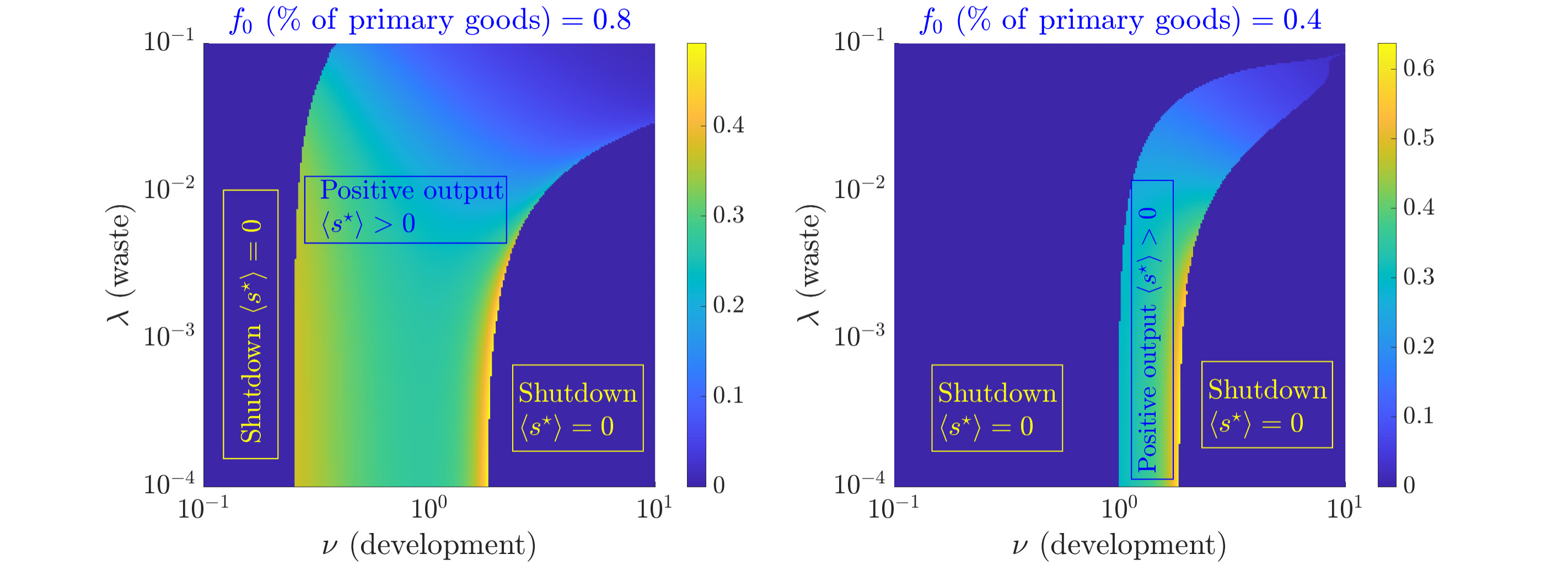}
\caption{Phase diagram of the model in the development ($\nu$) -- waste ($\lambda$)} plane, for different values of the fraction of primary goods $f_0$. Colors in the heatmaps correspond to equilibrium values of the optimal average scale of production $\langle s^\star \rangle$. As it can be seen, in both plots the ($\nu,\lambda$) plane features a region characterized by positive economic output ($\langle s^\star \rangle > 0$), which separates two regions of economic shutdown ($\langle s^\star \rangle = 0$).
\label{fig:nl}
\end{figure}

In Fig.~\ref{fig:nl} we report the average optimal scale of production $\langle s^\star \rangle$ (for which we derive a closed-form expression in Appendix~\ref{app:stat_prop}, see Eq.~\eqref{eq:avg_sstar}) at economic equilibrium in the $(\nu,\lambda)$ plane, both in the case of a resource-rich ($f_0 = 0.8$, left panel) and a resource-poor ($f_0 = 0.4$, right panels) economy. As it can be seen, as $\nu$ increases the economy undergoes two sharp transitions, delimiting a region characterized by positive economic output ($\langle s^\star \rangle > 0$)  from two regions in which no technology is active and the economy is effectively shut down ($\langle s^\star \rangle = 0$). The nature of such transitions will be explored in detail in a later Section. 

It is apparent from Fig.~\ref{fig:nl} that resource-rich economies have a much broader region of existence. The two regions of economic shutdown are therefore very different in magnitude depending on the value of $f_0$, but can nevertheless be characterized rather clearly. One of them corresponds roughly to the `south-east' quadrant of the $(\nu,\lambda)$ plane, meaning that no economy can simultaneously sustain high levels of development and low levels of waste. The other one roughly corresponds to a low-development `strip' in the $(\nu,\lambda)$ plane, i.e., at lower values of $\nu$ we find no economic activity, with the threshold being much lower for more resource-rich economies.

\begin{figure}[h!]
\centering
\includegraphics[width=\textwidth]{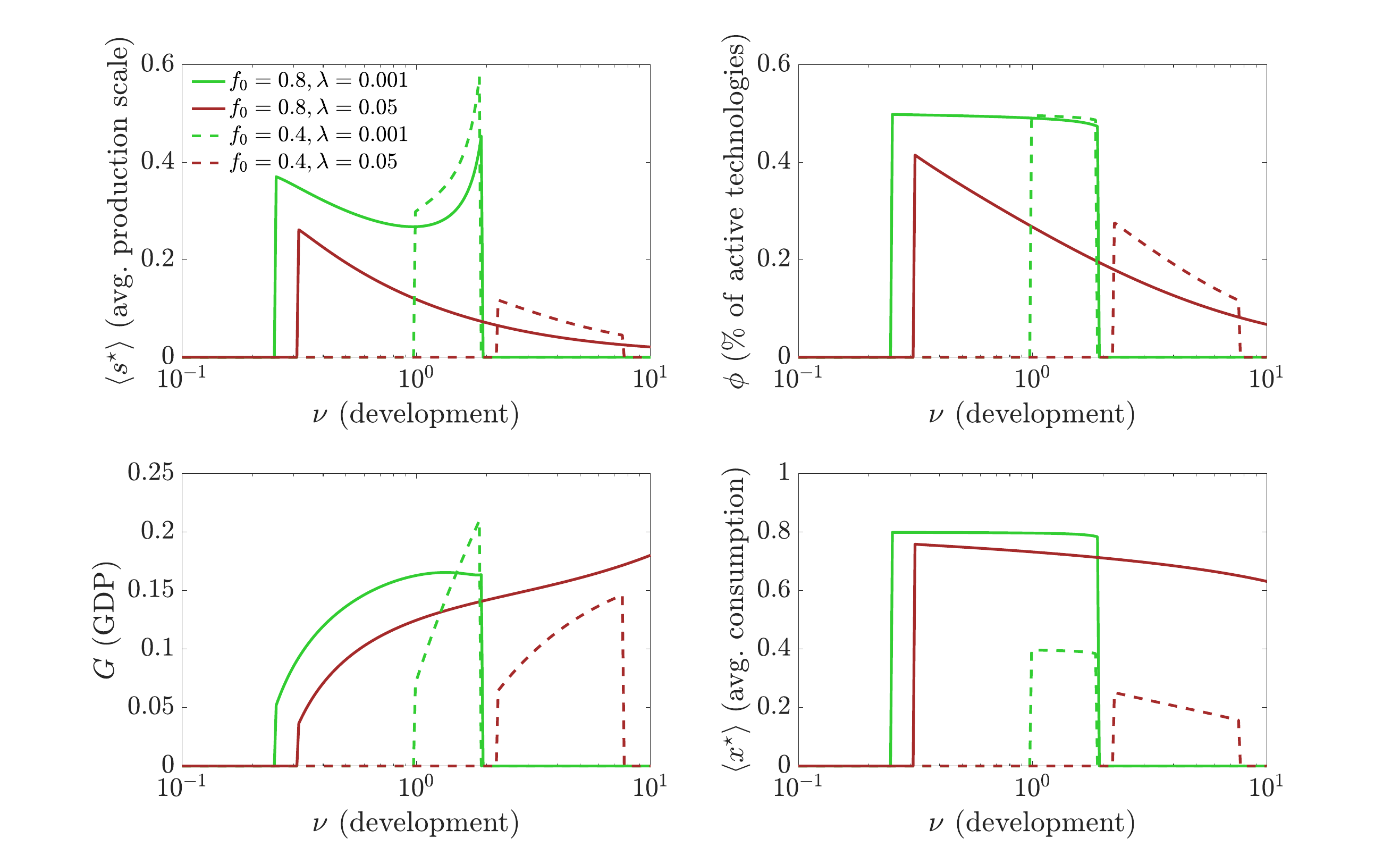}
\caption{Average scales of production $\langle s^\star \rangle$ (top left), fraction of active technologies $\phi$ (top right), GDP $G$ (bottom left) and average consumption $\langle x^c \rangle$ (bottom right) as a function of an economy's development level $\nu = N/C$ in the thermodynamic limit $N, C \rightarrow \infty$. Each quantity is obtained from the numerical solution of the saddle point equations in Eq.~\eqref{eq:sp}. In each plot, we report each quantity in a green ($\lambda = 0.001$, green lines) and brown ($\lambda =0.05$, brown lines) economy, both in a resource-rich ($f_0 = 0.8$, solid lines) and resource-poor ($f_0 = 0.4$, dashes lines) setting.}
\label{fig:curves}
\end{figure}

The plots in Fig.~\ref{fig:curves} provide some food for thought concerning an economy's path towards waste reduction, an obviously salient point in view of most real-world economies' efforts to curb emissions and transition towards `greener' industrial paradigms. In the top-left panel we show sample horizontal `slices' of the heatmaps in Fig.~\ref{fig:nl} corresponding to 
the average scale of production $\langle s^\star \rangle$ as a function of $\nu$ in four sample scenarios obtained combining the values $f_0 = 0.8$ (resource-rich economy) and $f_0 = 0.4$ (resource-poor economy) with $\lambda = 0.001$ (`green' economy) and $\lambda = 0.05$ (`brown' economy). Together with $\langle s^\star \rangle$, for the same combinations of values we also plot the average fraction of active firms/technologies $\phi = \lim_{N \rightarrow \infty} \sum_{n=1}^N \Theta(s_n^\star) / N$ (top-right panel), the economy's GDP $G$ (bottom-left panel), and the average consumption $\langle x^\star \rangle$ (bottom-right panel), see Appendix~\ref{app:stat_prop} for closed or semi-closed form expressions for such quantities.

Those plots highlight notable tradeoffs. First, from the top-left panel one can notice how green (green lines) vs brown (brown lines) economies behave differently as they develop, and as they approach the aforementioned phase transitions. Notably, the former experience a rather sharp increase in the average scale of production $\langle s^\star \rangle$ as they approach the highest level of development $\nu$ they can sustain before shutting down. Conversely, the scales of production in brown economies decrease monotonically as a function of $\nu$, and are systematically lower in value than in green economies with comparable resource availability. Similar considerations apply to the average fraction of active technologies $\phi$, which in green economies remains stable over the economy's region of existence, and systematically higher than in brown economies, where, again, we see a monotonic decline in economic activity as a function of $\nu$. 

At the same time, one can notice how brown economies can sustain higher levels of development compared to greener ones with the same resource availability, as their region of existence --- at comparable values of $f_0$ --- stretches out over higher values of $\nu$. On the one hand, this is rather unsurprising, as there is a somewhat obvious tradeoff between waste and economic development. On the other hand, it should be noted that, at development levels $\nu$ such that the two types of economies can coexist, greener ones have a systematic advantage, as they sustain higher scales of production, a larger fraction of active technologies, higher GDP and higher consumption levels.

\begin{figure}[h!]
\centering
\includegraphics[width=\textwidth]{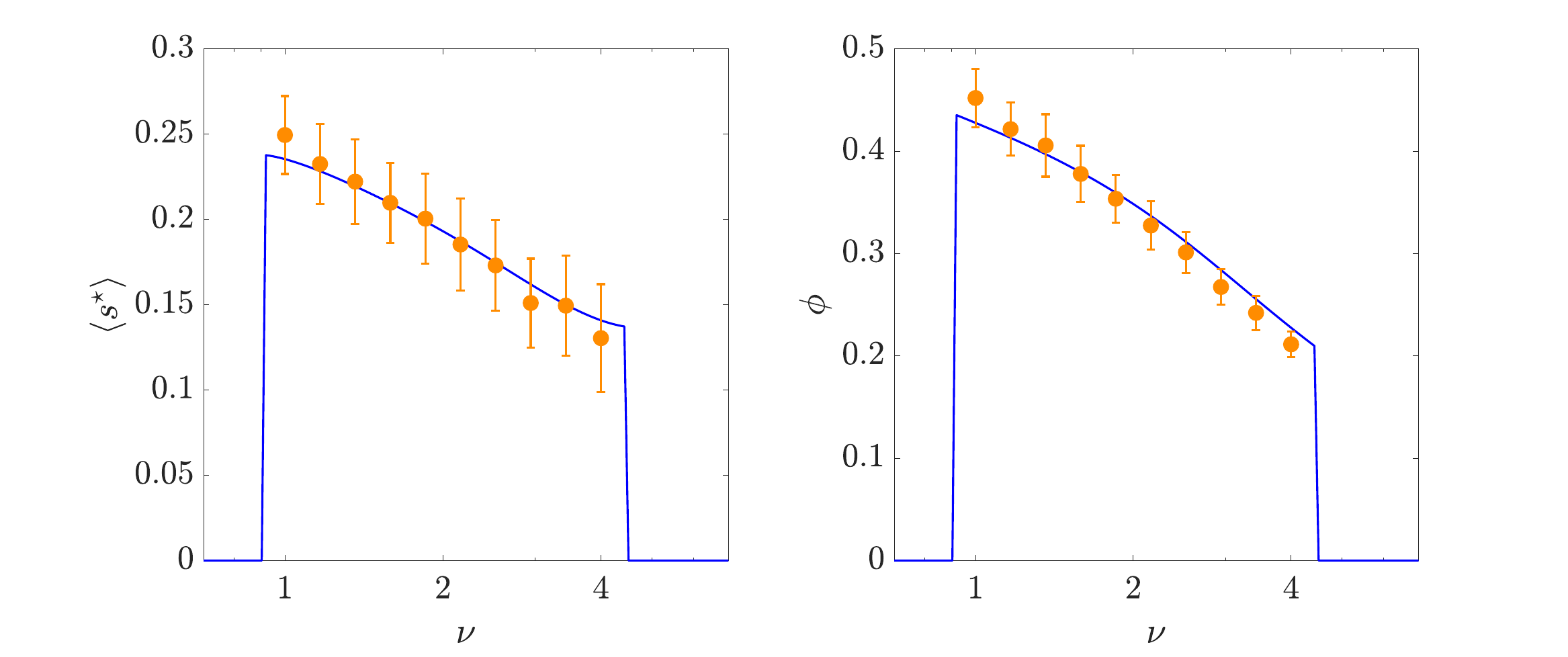}
\caption{Comparision between the average optimal scales of production (left) and fraction of active technologies (right) as a function of $\nu$ (with $f_0 = 0.5$ and $\lambda = 0.02$) as obtained in the thermodynamic limit from the solution of the saddle point equations~\eqref{eq:sp} (blue solid lines) and the numerical solution of the optimization problem in Eq.~\eqref{eq:opt_scales} (orange circles) for $N = 150$ and $C = N/\nu$ (rounded to the nearest integer). For each value of $\nu$, we generate $100$ independent realizations of the economy. The circles in the figures denote average values over such realizations, and the error bars represent the standard deviation.}
\label{fig:data}
\end{figure}

As a sanity check, we tested the validity of the analytical results we obtained from the (numerical) solution of the saddle point equations (see Eq.~\eqref{eq:sp} and Appendix~\ref{app:sp}) against the `brute force' solution of the problem in Eq.~\eqref{eq:opt_scales} for finite $N$ via numerical optimization. Sample results for $N = 150$ are presented in Fig.~\ref{fig:data}. As it can be seen, we find very good agreement between numerical simulations and theory.

\section{Nature of the phase transitions} \label{sec:pt}

In this Section, we provide some numerical evidence on the nature of the phase transitions illustrated in Fig.~\ref{fig:nl}. Their origins can be traced back to Eq.~\eqref{eq:xc}, which effectively identifies a set of constraints. Namely, the economy's production network matrix $q_n^c$ and the optimal scales of production at equilibrium $s_n^\star$ must guarantee that the production level of each good is a non-negative quantity, i.e., $(x^c)^\star \geq 0$. Clearly, at finite $N$ and $C$, this depends on the specific draw of $q_n^c$ from the probability density in Eq.~\eqref{eq:IO_pdf}. It may very well be that, for some `unlucky' draw of the input-output matrix, no production scale values exist to guarantee non-negative consumption levels for all goods, making the `shut down' solution $s_n^\star = 0$, $\forall \ n$ (and therefore $(x^c)^\star = 0$, $\forall \ n$) the only viable one. 

Remarkably, some analytical considerations can be made in order to characterize the above statistical process from a more formal point of view. To that end, let us introduce the following quantity:
 \begin{equation} \label{eq:vol}
V = \int_0^\infty \mathrm{d}\boldsymbol{s} \prod_{c=1}^C \Theta \left (x_0^c + \sum_{n=1}^N s_n q_n^c \right ) \ .
\end{equation}
In the above expression, we are integrating over all possible production scales---for fixed initial endowments and production matrix---a simple indicator function constructed as a product of Heaviside functions, i.e., $\Theta(x) = 1$ for $x \geq 0$ and $\Theta(x) = 0$ otherwise. Should our draw of the production matrix $q_n^c$ be particularly `unlucky', it will be impossible to find even a single vector $\boldsymbol{s}$ that prevents at least one of the $C$ Heaviside functions from becoming equal to zero, resulting in $V = 0$. Of course, we are not just interested in single instances of the production matrix, but rather on the average behavior of the quantity in Eq.~\eqref{eq:vol}, which effectively quantifies the `volume' in the space of production scales where economic activity can take place, in the thermodynamic limit $N, C \rightarrow \infty$, with $\nu = N/C$ fixed. Specifically, we are interested in determining under what conditions, and for which parameter values, we can expect the above volume to shrink to zero.

We numerically address the above problem by solving the following linear programming problem at finite $N$ and $C$ values
\begin{equation} \label{eq:lin_prog}
\max_{\boldsymbol{s}} \ \boldsymbol{s} \cdot \boldsymbol{\rho} \  \ \ \ \ \ \ \mathrm{subject \ to} \ \ \ \ 
\begin{cases}
      x_0^c + \sum_{n=1}^N s_n q_n^c \geq 0 & \forall \ c \\
      s_n \geq 0 & \forall \ n \ ,
    \end{cases}
\end{equation}
where $\rho$ is a random vector of positive numbers (e.g., $N$ i.i.d. draws from the uniform distribution in $[0, 1]$). Our focus in the above problem is not on its solution, but rather on the \emph{existence} of a non-trivial solution, hence our use of a random vector. In other words, the above linear programming problem looks for solutions within the polytope identified by its constraints. Should non-trivial solutions exist, then the constraints do not shrink the volume of possible solutions in Eq.~\eqref{eq:vol} to just the origin ($s_n = 0$, $\forall \ n$). 

\begin{figure}[h!]
\centering
\includegraphics[width=0.5\textwidth]{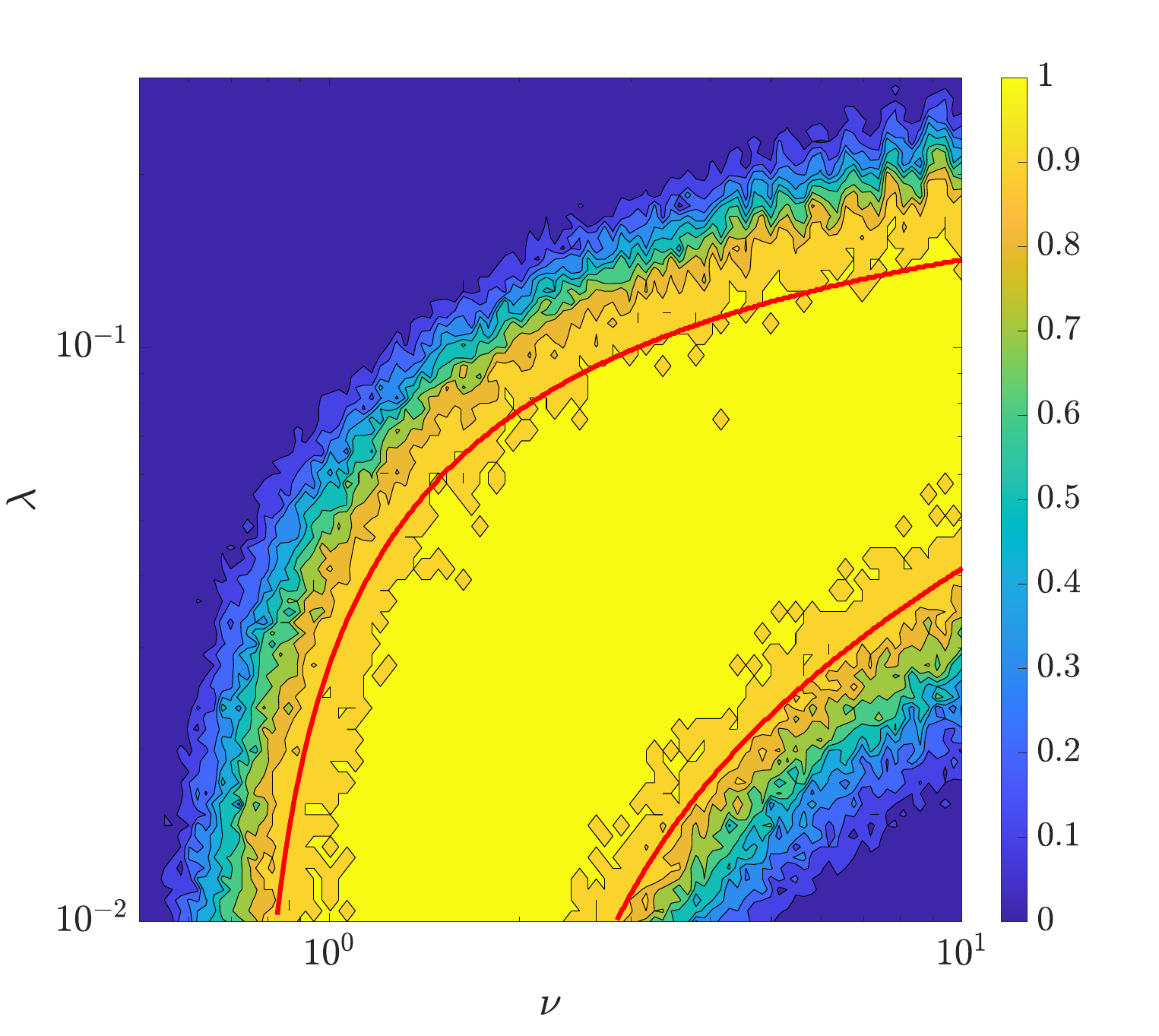}
\caption{Comparison between the critical lines --- obtained as the values of $\nu$, $\lambda$, and $f_0$ for which the saddle point equations~\eqref{eq:sp} cease to have solutions --- separating the regions of economic shutdown (shown in blue in Fig.~\ref{fig:nl}) from those where there is economic activity (red solid lines) and the regions where solutions to the linear programming problem in Eq.~\eqref{eq:lin_prog} can be found. Regions in yellow correspond to parameter combinations for which a solution to Eq.~\eqref{eq:lin_prog} can be found with probability $\approx 1$, regardless of the specific choice of the random vector $\boldsymbol{\rho}$. Conversely, region in blue correspond to parameter combinations for which Eq.~\eqref{eq:lin_prog} has no solutions.}
\label{fig:transition}
\end{figure}

In Fig.~\ref{fig:transition} we plot the fraction of linear programming problems as in Eq.~\eqref{eq:lin_prog} for which we obtain a non-trivial solution ($s_n > 0$ for at least some $n$) as a function of $\nu$ and $\lambda$, with $f_0 = 0.5$ and $N = 100$. For each combination of $\nu$ and $\lambda$, we generate $50$ independent input-output matrices, and solve Eq.~\eqref{eq:lin_prog} for each. The red lines denote the critical lines --- obtained from the numerical solution of the saddle point equations --- separating the regions of economic activity and economic shut down. As it can be seen, despite the relatively small size of the systems being simulated here, we find good agreement with the thermodynamic limit results.

\begin{figure}[h!]
\centering
\includegraphics[width=0.7\textwidth]{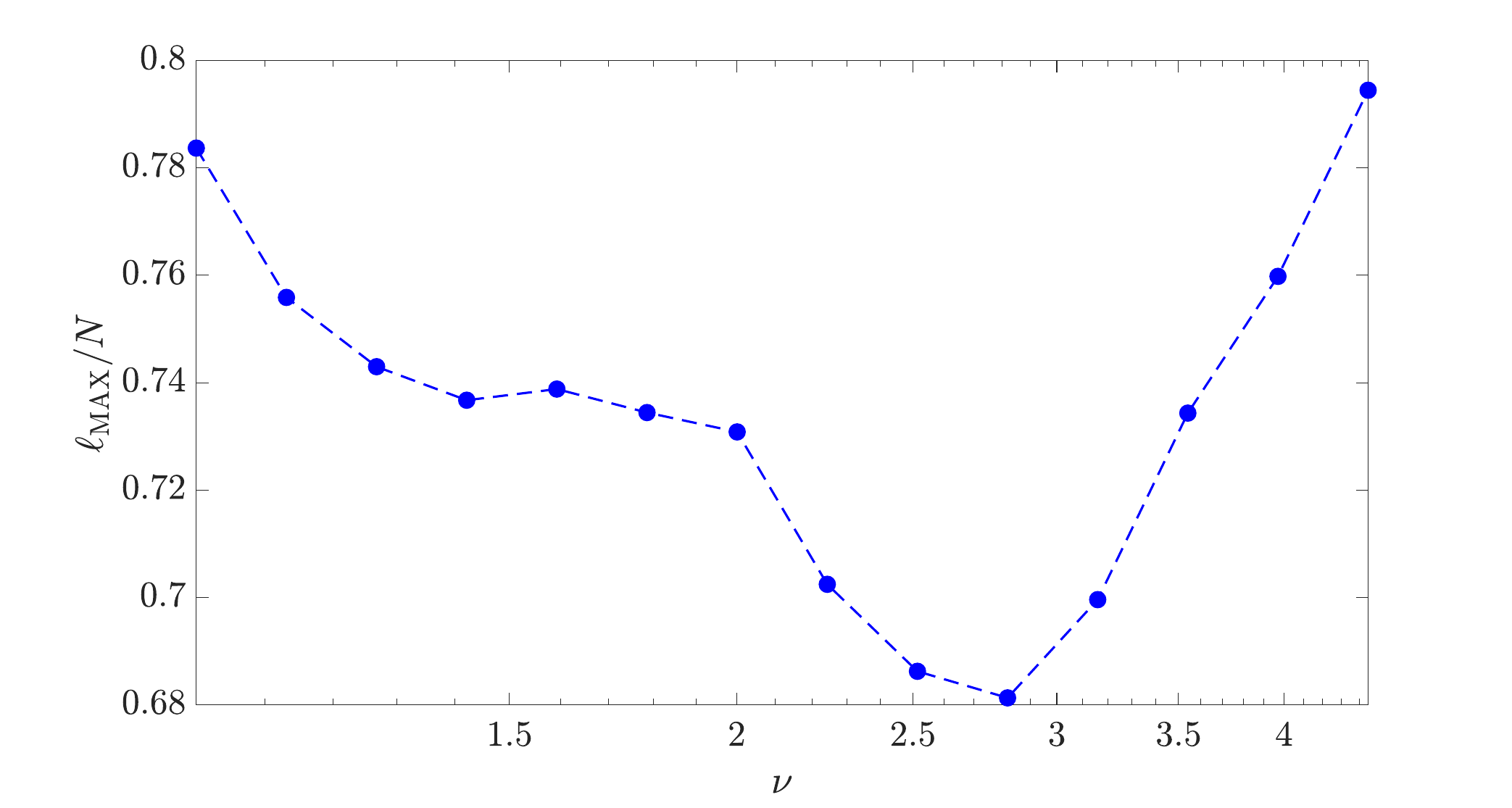}
\caption{Average leading eigenvalue $\ell_\mathrm{MAX}$ (normalized by $N$) of the correlation matrix of different solutions $\boldsymbol{s}^\star$ of the linear programming problem in Eq.~\eqref{eq:lin_prog} obtained for different random choices of $\boldsymbol{\rho}$. We choose $N = 150$, and fix $C = N / \nu $ accordingly (rounding up to the nearest integer). For each value of $\nu$, we generate 500 different economies (i.e., different draws of $x_0^c$ and $q_n^c$), using $f_0 = 0.5$ and $\lambda = 0.02$. For each of those economies, we generate 100 different solutions of the linear programming problem, and use them to form the aforementioned correlation matrix.}
\label{fig:PCA}
\end{figure}

We further corroborate the above picture by solving the linear programming problem in Eq.~\eqref{eq:lin_prog} --- at fixed $x_0^c$ and $q_n^c$, i.e., for the same economy --- for several different random choices of the vector $\rho$. For each economy, we build the correlation matrix of the solutions $\boldsymbol{s}^\star$ obtained in such a way, and in Fig.~\ref{fig:PCA} we report the behavior of their leading eigenvalue $\ell_\mathrm{MAX}$, averaged over several different economies. As known from principal component analysis~\cite{livan2018random}, values $\ell_\mathrm{MAX}/N \rightarrow 1$ indicate very strong correlations, which in our case indicate that the solutions $\boldsymbol{s}^\star$ obtained for different random draws of $\boldsymbol{\rho}$ are exceedingly similar to each other, which, in turn, indicates that the volume in Eq.~\eqref{eq:lin_prog} must be shrinking. As it can be seen from Fig.~\ref{fig:PCA}, $\ell_\mathrm{MAX}/N$ displays a rather clear U-shaped behavior as a function of $\nu$, getting closer to one as $\nu$ approaches its two critical values ($\nu \approx 0.92$ and $\nu \approx 4.43$, respectively) for the parameters used in the simulation ($f_0 = 0.5$ and $\lambda = 0.02$).  

Put together, the above numerical evidence is very much reminiscent of phase transitions in high-dimensional random geometries such as those found, e.g., in models of neural networks~\cite{gardner1988space} and financial markets~\cite{bardoscia2012financial}, which have been characterized systematically in~\cite{donoho2009observed}.

\section{Discussion}

In this work, we developed a stylized model of a complex economy to analyze the trade-offs associated with the green transition, using techniques from the Statistical Mechanics of disordered systems. By promoting the parameters of a General Equilibrium Theory (GET) model to random variables, we characterized the typical, self-averaging properties of large random economies under environmental constraints, akin to those that modern developed economies are increasingly expected to face under the so called green transition. A central feature of our model is the explicit inclusion of a waste variable, which serves as a proxy for emissions or other undesirable byproducts of economic activity.

Our model necessarily makes a number of simplifying assumptions, and it inherits from GET the limitation of not capturing temporal dynamics, as GET focuses on static equilibria. Nonetheless, these equilibria can be interpreted in a `meta-stable' sense---that is, as economic states that may persist for extended periods before transitioning to a new equilibrium due to shifts in the economy's fundamental parameters. Despite its limitations, our model captures key qualitative features of modern economies under environmental pressure. The most notable outcome is the emergence of a double phase transition in the $(\nu,\lambda)$ plane, where $\nu = N/C$ denotes the ratio of technologies to goods (which proxies economic development) and $\lambda$ characterizes waste intensity. Economic activity is confined to a bounded `feasible' region, surrounded by two shutdown regimes. 

The first, at low $\nu$, corresponds to a \textit{technology-scarcity regime}: the economy cannot sustain positive production because there are too few viable technological combinations to meet the basic non-negativity constraints on consumption (Eq.~\eqref{eq:xc}). This regime is reminiscent of underdevelopment traps~\cite{rosenstein1943problems}, where lack of industrial diversity prevents economic takeoff. The second, at high $\nu$, is a \textit{waste-constrained regime}: while the economy possesses many technologies, the cumulative waste generated by activating them shrinks the feasible production space until it collapses. This reflects the tension between technological abundance and environmental constraints, and parallels the notion of planetary boundaries in sustainability studies~\cite{rockstrom2009planetary}.

Our results also highlight systematic differences between `brown' and `green' economies. Brown economies (high $\lambda$) can, in principle, access higher $\nu$ values, corresponding to a larger technological base, but they activate only a small fraction of their technologies as development progresses. This leads to lower production scales, GDP, and average consumption. Green economies (low $\lambda$), by contrast, operate with fewer technologies per good but utilize a higher fraction of them, achieving greater economic welfare across most of their feasible region. However, as green economies approach the high-$\nu$ boundary, they become vulnerable to abrupt shutdowns if environmental constraints tighten or if production scales are pushed too aggressively. 

These findings offer an interpretation in standard economic terms. In classical GET, the production possibility frontier (PPF)---the boundary between what can and cannot be produced given a set of resources and technologies~\cite{mas1995microeconomic}---is smooth and convex. In our statistical mechanical formulation, the effective PPF can fragment and ultimately vanish as complexity and environmental pressure interact. The `volume collapse' of feasible allocations that drives the double phase transition is analogous to the shrinking of solution spaces in high-dimensional combinatorial problems, a phenomenon well documented in the study of algorithmic phase transitions~\cite{monasson1999determining,mezard2002analytic}. From an economic perspective, this reflects the increasing difficulty of coordinating production in highly complex, environmentally constrained economies, and resonates with concerns about structural fragility in global value chains~\cite{carvalho2021supply,pichler2023building}.

Overall, our analysis provides a stylized yet insightful framework for thinking about the structural risks of the green transition. It shows that low-waste technological diffusion can initially enhance economic welfare by enabling higher activation of productive technologies. However, the feasible space for economic activity remains bounded, and rapid growth in a tightening environmental regime may lead to sudden collapses in output. These insights emphasize the importance of managing both technological development and environmental capacity in a coordinated manner to ensure a smooth transition to sustainable production paradigms.

\appendix

\section{Statistical mechanical formulation} \label{app:statmech}

We are interested in solving the optimization problem in Eq.~\eqref{eq:opt_scales} as the economy grows large, i.e., when $N, C \rightarrow \infty$, with $\nu = N/C$ fixed. Thus, we are effectively interested in calculating the maxima of the utility function $U_N$ in the following limit:
\begin{equation} \label{eq:opt_prob}
\lim_{N \rightarrow \infty} \frac{1}{N} \max_{\boldsymbol{s} \geq 0} \ U_N(\boldsymbol{s} | \boldsymbol{q}, \boldsymbol{\epsilon}, \boldsymbol{x}_0) =
\lim_{N \rightarrow \infty} \lim_{\beta \rightarrow \infty} \frac{1}{\beta N} \log Z_N (\beta | \boldsymbol{q}, \boldsymbol{\epsilon}, \boldsymbol{x}_0) \ ,
\end{equation}
where we have introduced the following function. 
\begin{equation*}
Z_N (\beta | \boldsymbol{q}, \boldsymbol{\epsilon}, \boldsymbol{x}_0) = \int \de \boldsymbol{s} \ \mathrm{e}^{\beta U_N(\boldsymbol{s} | \boldsymbol{q}, \boldsymbol{\epsilon}, \boldsymbol{x}_0)} \ .
\end{equation*}
With this position, the optimization problem in Eq.~\eqref{eq:opt_prob} is effectively converted into a steepest descent search, as the limit $\beta \rightarrow \infty$ selects the values of the utility function $U_N$ that dominate the integral defining $Z_N$, which in turn provides us with the desired result. In Statistical Mechanics, $Z_N$ obviously mirrors a canonical partition function (and its logarithm mirrors a free energy), with the parameter $\beta$ playing the role of an inverse temperature. According to this interpretation, Eq.~\eqref{eq:opt_prob} informs us that the optimization problem we are interested in is fully equivalent to computing the zero temperature limit of the free energy of the system described by the partition function $Z_N$.

Reasonably, one can expect the maxima of the function $U_N$ to grow with the system's size $N$, so that the quantity $U_N/N$ in Eq.~\eqref{eq:opt_prob} will remain finite (intensive, in statistical mechanical parlance) in the $N \rightarrow \infty$ limit. Thus, our position is that, for a reasonable choice of the utility function, the maxima of $U_N/N$ will cease to depend on the economy's details as it grows large, i.e., they will obey
\begin{equation}
\lim_{N \rightarrow \infty} \frac{1}{N} \max_{\boldsymbol{s} \geq 0} \ U_N(\boldsymbol{s} | \boldsymbol{q}, \boldsymbol{\epsilon}, \boldsymbol{x}_0) =
\lim_{N \rightarrow \infty} \frac{1}{N} \left \langle \max_{\boldsymbol{s} \geq 0} \ U_N(\boldsymbol{s} | \boldsymbol{q}, \boldsymbol{\epsilon}, \boldsymbol{x}_0) \right \rangle_{\boldsymbol{q}, \boldsymbol{\epsilon}, \boldsymbol{x}_0} = 
\lim_{N \rightarrow \infty} \lim_{\beta \rightarrow \infty} \frac{1}{\beta N} \Big \langle \log Z_N(\beta | \boldsymbol{q}, \boldsymbol{\epsilon}, \boldsymbol{x}_0) \Big \rangle_{\boldsymbol{q}, \boldsymbol{\epsilon}, \boldsymbol{x}_0} \ .
\end{equation}
From a practical standpoint, the above expression looks rather inconvenient, as it requires averaging over the logarithm of the partition function $Z_N$. Such a rather cumbersome task is usually circumvented via the so-called replica trick, which effectively lets us compute the average of the logarithm as the logarithm of an average. This is achieved by first converting the average over $\log Z_N$ into an average over $r$ replicas of the system---that is, the factorization of $r$ identical partition functions---through the formal identity
\begin{equation*}
\langle \log Z_N \rangle = \lim_{r \rightarrow 0} \frac{\langle Z_N^r \rangle - 1}{r} \ .
\end{equation*}
When using replicas, $r$ is assumed to be an integer number throughout all calculations, and restored to be a real number only at the end by taking the above limit.

All in all, the optimization problem in Eq.~\eqref{eq:opt_prob} can be solved as
\begin{equation} \label{eq:apphtheta}
\lim_{N \rightarrow \infty} \frac{1}{N} \left \langle \max_{\boldsymbol{s} \geq 0} \ U_N(\boldsymbol{s} | \boldsymbol{q}, \boldsymbol{\epsilon}, \boldsymbol{x}_0) \right \rangle_{\boldsymbol{q}, \boldsymbol{\epsilon}, \boldsymbol{x}_0} = 
\lim_{N \rightarrow \infty} \lim_{\beta \rightarrow \infty} \lim_{r \rightarrow 0} \frac{1}{\beta N r} \log \Big \langle Z_N^r (\beta | \boldsymbol{q}, \boldsymbol{\epsilon}, \boldsymbol{x}_0) \Big \rangle_{\boldsymbol{q}, \boldsymbol{\epsilon}, \boldsymbol{x}_0} = h(\boldsymbol{\theta}^\star) \ ,
\end{equation}
where the last equality comes from the fact that the replica trick allows to identify a small set of \emph{order parameters} $\boldsymbol{\theta} = (\theta_1, \ldots, \theta_k)$ such that
\begin{equation} \label{eq:replicated_Z}
\Big \langle Z_N^r (\beta | \boldsymbol{q}, \boldsymbol{\epsilon}, \boldsymbol{x}_0) \Big \rangle_{\boldsymbol{q}, \boldsymbol{\epsilon}, \boldsymbol{x}_0} \approx
\int \de \boldsymbol{\theta} \ \mathrm{e}^{\beta N r h(\boldsymbol{\theta})}
\end{equation}
for a suitable function $h$, and the optimal order parameter values in Eq.~\eqref{eq:apphtheta} are found by solving the set of saddle point equations
\begin{equation*}
\left ( \frac{\partial h}{\partial \theta_i} \right )_{\theta_i = \theta_i^\star} = 0 \ \qquad \ i = 1, \ldots, k
\end{equation*}
which extract the maximum of the function in the exponential in the above integral.

In Appendix~\ref{app:sp} we provide a detailed derivation of the solution to the above optimization problem, including the identification of the corresponding order parameters. Let us emphasize here that identifying a system's order parameters in a replica trick calculation is more of an art than a science, as there is no straightforward `algorithmic' procedure to do that. However, identifying a system's order parameters simplifies its description quite considerably. As a matter of fact, one should not lose sight of the fact that, when taking the thermodynamic limit, one is dealing with infinitely many variables (infinitely many firms and goods, in our case). Yet, when the order parameters have been identified, one ends up---through Eq.~\eqref{eq:apphtheta}---with a \emph{complete description} of the system based on just $k$ variables, where $k$ is usually a small number (in our case, we will see that $k=6$).

\section{The optimization problem} \label{app:sp}

The starting point of our derivation is the replicated partition function appearing in Eq.~\eqref{eq:replicated_Z}, which we can write as 
\begin{equation*}
Z_N^r (\beta | \boldsymbol{q}, \boldsymbol{\epsilon}, \boldsymbol{x}_0) = \int_{0}^{\infty}\prod_{a=1}^{r}\de \boldsymbol{x}_a\int_{0}^{\infty} \prod_{a=1}^{r} \de \boldsymbol{s}_a \mathrm{e}^{\beta \sum_{a=1}^{r}U(\boldsymbol{x}_a)}\prod_{a=1}^{r}\prod_{c=1}^{C}\delta \left (x^c-x_0^c-\sum_{n=1}^{N}s_{n,a}q_n^c \right) \ ,
\end{equation*}
where the delta functions implement the constraints in Eq.~\eqref{eq:xc}. As per Eq.~\eqref{eq:apphtheta}, we now need to average the above expression with respect to the PDFs in Eqs.~\eqref{eq:waste_pdf},~\eqref{eq:IO_pdf} and~\eqref{eq:f0_pdf}. Using the integral representation of the delta function $\delta(x) = (2\pi)^{-1} \int_{-\infty}^\infty \de \hat{k} \ \mathrm{e}^{\mathrm{i} \hat{k} x}$, we can express the average of the delta functions appearing in the partition function as
\begin{eqnarray*}
&& \left \langle \prod_{a=1}^{r}\prod_{c=1}^{C}\delta \left (x^c-x_0^c-\sum_{n=1}^{N}s_{n,a}q_n^c \right) \right \rangle_{\boldsymbol{q}, \boldsymbol{\epsilon}} = \\ \nonumber
&& \int_{-\infty}^\infty \prod_{n=1}^N \frac{\de \hat{z}_n}{2\pi} \int_{-\infty}^\infty \prod_{a=1}^r \frac{\de \hat{\boldsymbol{x}}_a}{2\pi} \exp\left[\mathrm{i} \sum_{a=1}^{r}\sum_{c=1}^{C}\hat{x}_a^c\left(x_a^c-x_0^c\right)-\frac{1}{2C}\sum_{n=1}^{N}\sum_{c=1}^{C}\left(\hat{z}_n-\sum_{a=1}^{r}s_{n,a}\hat{x}_a^c\right)^2-\frac{\lambda}{2}\left(\sum_{n=1}^{N}\hat{z}_n\right)^2+\mathrm{i}\epsilon_0\sum_{n=1}^{N}\hat{z}_n\right] \ ,
\end{eqnarray*}
which leads us to the following expression
\begin{eqnarray*}
\langle Z_N^r (\beta | \boldsymbol{q}, \boldsymbol{\epsilon}, \boldsymbol{x}_0) \rangle_{\boldsymbol{q}, \boldsymbol{\epsilon}} &=&
\left [\sqrt{2\pi\left(\lambda+1\right)} \exp\left(\frac{\epsilon_0^2}{2\left(\lambda+1\right)}\right) \right ]^N
\left \langle \int_{-\infty}^\infty \prod_{n=1}^N \frac{\de \hat{z}_n}{2\pi} \int_{-\infty}^\infty \prod_{a=1}^r \frac{\de \hat{\boldsymbol{x}}_a}{2\pi} \int_0^\infty \prod_{a=1}^r \de \boldsymbol{x}_a \int_0^\infty \prod_{a=1}^r \de \boldsymbol{s}_a \right. \nonumber \\
&&\left. \times \exp\left[\beta\sum_{a=1}^{r} U_N \left(\boldsymbol{x}_a\right) + \mathrm{i} \sum_{a=1}^{r}\sum_{c=1}^{C}\hat{x}_a^c\left(x_a^c-x_0^c\right)\right] \right. \nonumber \\
&&\left. \times  \exp\left[-\frac{1}{2C}\sum_{n=1}^{N}\sum_{c=1}^{C}\left(\hat{z}_n-\sum_{a=1}^{r}s_{n,a}\hat{x}_a^c\right)^2+ \mathrm{i}\sqrt{\lambda}w\sum_{n=1}^{N}\hat{z_n}+\mathrm{i}
\epsilon_0\sum_{n=1}^{N}\hat{z}_n \right] \right \rangle_w \ .
\end{eqnarray*}
In the above expression, the average $\langle \cdots \rangle_w$ comes from the Hubbard-Stratonovich transformation
\begin{equation*}
\exp \left (- \frac{a x^2}{2} \right) = \frac{1}{\sqrt{2 \pi}} \int_{-\infty}^\infty \de w  \exp \left (-\frac{w^2}{2} + \mathrm{i} \sqrt{a} x w  \right ) = \Big \langle \exp \left ( \mathrm{i} \sqrt{a} x w  \right ) \Big \rangle_w \  ,
\end{equation*}
which linearizes the term proportional to $\hat{z}^2$ at the cost of introducing an auxiliary Gaussian standard variable $w \sim \mathcal{N}(0,1)$. The prefactor $\left [\sqrt{2\pi\left(\lambda+1\right)} \exp\left( \epsilon_0^2 / 2\left(\lambda+1\right) \right) \right ]^N$ comes from integrating out the denominator in Eq.~\eqref{eq:IO_pdf} with respect to $\epsilon$ (i.e., with respect to the pdf in Eq.~\eqref{eq:waste_pdf}).

We can now introduce the following variables
\begin{equation*}
\omega_{ab}=\frac{1}{N}\sum_{n=1}^{N}s_{n,a}s_{n,b} \qquad k_a=\frac{1}{N}\sum_{n=1}^{N}\hat{z}_n s_{n,a} \ ,
\end{equation*}
the first one of which denotes the overlap between different replicas, and the following identities
\begin{eqnarray*}
    1 &= & N \int_{-\infty}^\infty \de \omega_{ab} \ \delta\left(N\omega_{ab}-\sum_{n=1}^{N}s_{n,a}s_{n,b}\right) = N \int_{-\infty}^\infty \frac{\de \omega_{ab} \de \hat{\omega}_{ab}}{2\pi \mathrm{i}}\exp\left[\hat{\omega}_{ab}\left(N\omega_{ab}-\sum_{n=1}^{N}s_{n,a}s_{n,b}\right)\right] \\ \nonumber
   1 & =& N \int_{-\infty}^\infty \de k_a \ \delta \left(N k_a-\sum_{n=1}^{N}\hat{z}_n s_{n,a}\right) = N \int_{-\infty}^\infty \frac{\de k_a \de \hat{k}_a}{2\pi\mathrm{i}}\exp\left[\hat{k}_a\left(Nk_a-\sum_{n=1}^{N}\hat{z}_n s_{n,a}\right)\right] \ ,
\end{eqnarray*}
thanks to which we can express the averaged replicated partition function as
\begin{equation} \label{eq:avg_replicated_Z}
\langle Z_N^r (\beta | \boldsymbol{q}, \boldsymbol{\epsilon}, \boldsymbol{x}_0) \rangle_{\boldsymbol{q}, \boldsymbol{\epsilon}, \boldsymbol{x}_0} =\int_{-\infty}^\infty \prod_{a,b=1}^{r}\frac{\de \omega_{ab} \de \hat{\omega}_{ab}}{4\pi \mathrm{i}/N} \int_{-\infty}^\infty \prod_{a=1}^{r}\frac{\de k_a \de \hat{k}_a}{2\pi \mathrm{i}/N}\exp\left[N h\left(\{\omega_{ab}\},\{\hat{\omega}_{ab}\},\{k_a\},\{\hat{k}_a\}\right)\right] \ .
\end{equation}
The function $h$ in the above expression reads
\begin{equation*}
h\left(\{\omega_{ab}\},\{\hat{\omega}_{ab}\},\{k_a\},\{\hat{k}_a\}\right)=g_1\left(\{\omega_{ab}\},\{\hat{\omega}_{ab}\},\{k_a\},\{\hat{k}_a\}\right)+g_2\left(\{\hat{\omega}_{ab}\},\{\hat{k}_a\}\right)+g_3\left(\{\omega_{ab}\},\{k_a\}\right) \ , 
\end{equation*} 
where 
\begin{eqnarray*}
g_1 &=& -\frac{1}{2}\sum_{a,b=1}^{r}\hat{\omega}_{ab}\omega_{ab}-\sum_{a=1}^{r}\hat{k}_a k_a \\ \nonumber
g_2 &=& \log\int_{-\infty}^{\infty}\frac{\de w}{\sqrt{2\pi}}\int_{-\infty}^{\infty}\frac{\de\hat{z}}{2\pi}\int_{0}^{\infty}\prod_{a=1}^{r}\de s_a\exp\left[\frac{1}{2}\sum_{a,b=1}^{r}\hat{\omega}_{ab}s_a s_b+\hat{z}\sum_{a=1}^{r}\hat{k}_as_a-\frac{\hat{z}^2}{2}+ \mathrm{i} \sqrt{\lambda}\hat{z}w-\frac{w^2}{2}+ \mathrm{i} \epsilon_0\hat{z}\right] \\ \nonumber
&+ & \log \left [\sqrt{2\pi\left(\lambda+1\right)} \exp\left(\frac{\epsilon_0^2}{2\left(\lambda+1\right)}\right) \right ] \\ \nonumber
g_3 &=& \frac{1}{N}\sum_{c=1}^{C}\log\int_{-\infty}^{\infty}\prod_{a=1}^{r}\frac{\de\hat{x}_a}{2\pi}\int_{0}^{\infty}\prod_{a=1}^{r}\de x_a\exp\left[\beta\sum_{a=1}^{r} U_N \left(x_a\right)+ \mathrm{i} \sum_{a=1}^{r}\hat{x}_a\left(x_a-x_0^c\right)-\frac{\nu}{2}\sum_{a,b=1}^{r}\hat{x}_a\hat{x}_b\omega_{ab}+\nu \sum_{a=1}^{r}\hat{x}_ak_a\right] \ .
\end{eqnarray*}

In the thermodynamic limit, the leading contributions to the integrals in Eq.~\eqref{eq:avg_replicated_Z} will come from $h\left(\{\omega^\star_{ab}\},\{\hat{\omega}^\star_{ab}\},\{k^\star_a\},\{\hat{k}^\star_a\}\right)$, where the star denotes the values at which $h$ reaches its maximum. Under a replica symmetric ansatz, we can expect solutions of the form
\begin{eqnarray*}
\omega_{ab}^\star &=& \Omega\delta_{ab}+\omega\left(1-\delta_{ab}\right) \\ \nonumber
\hat{\omega}_{ab}^\star &=& \hat{\Omega}\delta_{ab}+\hat{\omega}\left(1-\delta_{ab}\right) \\ \nonumber
k^\star_a &=& k \\ \nonumber
\hat{k}_a^\star&=& \hat{k} \ ,
\end{eqnarray*} 
which let us readily compute expressions for the functions $g_1$, $g_2$, and $g_3$ in the limit $r \rightarrow 0$

It is easy to find an analytic expression of the function $g_1$, $g_2$ and $g_3$ in terms of r and to perform the limit $r\rightarrow0$ (see Eq.~\eqref{eq:htheta}):
\begin{eqnarray*}
\lim_{r \to 0}\frac{g_1}{r}&=& -\frac{\hat{\Omega}\Omega-\hat{\omega}\omega}{2}-\hat{k}k \\ \nonumber
\lim_{r \to 0}\frac{g_2}{r} &=& \left \langle \log\int_{0}^{\infty} \de s \ \mathrm{e}^{\beta V(s,t)}\right \rangle_t \\ \nonumber
\lim_{r \to 0}\frac{g_3}{r} &=& \frac{1}{\nu}\left \langle \log\int_{0}^{\infty} \de x \ \mathrm{e}^{\beta W(x,x_0,t)}\right \rangle_{t,x_0} \ ,
\end{eqnarray*}
where, again, $t \sim \mathcal{N}(0,1)$, and the functions $\beta V(s,t)$ and $\beta W (x,x_0,t)$ read
\begin{eqnarray} \label{eq:VW_func}
\beta V(s,t) &=& \frac{\hat{\Omega}-\hat{\omega}}{2}s^2+st \sqrt {\hat{\omega}+\frac{ \hat{k}^2}{\lambda+1} } +\mathrm{i}\frac{\hat{k}s}{\lambda+1}\epsilon_0 \\ \nonumber
\beta W(x,x_0,t) &=& \beta u\left(x\right)-\frac{\left(x-x_0+ \sqrt{\nu \omega} t- \mathrm{i} \nu  k\right)^2}{2\nu \left(\Omega-\omega\right)}-\frac{1}{2}\log\left[2\pi \nu \left(\Omega-\omega\right)\right] \ .
\end{eqnarray}

In order to proceed with the limit $\beta \rightarrow \infty$ in Eq.~\eqref{eq:htheta}, we can introduce the following parameters
\begin{equation*}
\chi=\nu \beta\left(\Omega-\omega\right) \qquad \hat{\chi}=-\frac{\hat{\Omega}-\hat{\omega}}{\beta} \qquad \kappa=-\mathrm{i}\nu\left(1+\lambda\right) k \qquad \hat{\kappa}=\frac{\mathrm{i}\hat{k}}{\left(1+\lambda\right)\beta} \qquad \hat{\gamma}=\frac{\hat{\omega}}{\beta^2} \ ,
\end{equation*}
rewrite $h$ as
\begin{equation*}
h\left(\Omega,\kappa,\hat{\kappa},\hat{\gamma},\chi,\hat{\chi}\right)=\frac{1}{2}\left(\Omega\hat{\chi}-\frac{\hat{\gamma}\chi}{\nu}\right)-\frac{\hat{\kappa}\kappa}{\nu}+\frac{1}{\beta}\left \langle \log\int_{0}^{\infty} \de s \ \mathrm{e}^{\beta V(s,t)}\right \rangle_t 
+\frac{1}{\nu\beta}\left \langle \log\int_{0}^{\infty} \de x \ \mathrm{e}^{\beta W(x,x_0,t)}\right \rangle_{t,x_0} \ ,
\end{equation*}
and the functions in Eq.~\eqref{eq:VW_func} as
\begin{eqnarray*}
V(s,t) &=& -\frac{\hat{\chi}}{2}s^2+st\sqrt{\hat{\gamma}-\left(1+\lambda\right)\hat{\kappa}^2}+\hat{\kappa}\epsilon_0s \\ \nonumber
W(x,x_0,t) &=& u\left(x\right)-\frac{1}{2\chi} \left(x-x_0+ t \sqrt{\nu\Omega} +\frac{\kappa}{1+\lambda}\right)^2-\frac{1}{2\beta}\log\left(\frac{2\pi\chi}{\beta}\right) \ .
\end{eqnarray*}
As long as $\chi$ remains finite, the third terms in both $V(s,t)$ and $W(x,x_0,t)$ become negligible in the $\beta \rightarrow \infty$ limit. Thus, in such a limit we can express the function $h$ as
\begin{eqnarray} \label{eq:heq}
h\left(\Omega,\kappa,\hat{\kappa},\hat{\gamma},\chi,\hat{\chi}\right) &=& \frac{1}{2}\left(\Omega\hat{\chi}-\frac{\hat{\gamma}\chi}{\nu}\right)-\frac{\hat{\kappa}\kappa}{\nu}+\left \langle \max_{s\ge0}\left[-\frac{\hat{\chi}}{2}s^2+st\sqrt{\hat{\gamma}-\left(1+\lambda\right)\hat{\kappa}^2}+\hat{\kappa}\epsilon_0s\right]\right \rangle_t \\ \nonumber
&+& \frac{1}{\nu} \left \langle \max_{x\ge0} \left [ u(x) - \frac{1}{2\chi} \left(x-x_0+ t \sqrt{\nu\Omega} +\frac{\kappa}{1+\lambda}\right)^2 \right ] \right \rangle_{t,x_0}
\end{eqnarray}
Indicating with $s^\star$ and $x^\star$ the values at which the last two terms in the above expressions reach their maxima (whose statistical properties we characterize in the next Section), we get
\begin{eqnarray} \label{eq:h_app}
h\left(\Omega,\kappa,\hat{\kappa},\hat{\gamma},\chi,\hat{\chi}\right) &=& \frac{1}{2}\left(\Omega\hat{\chi}-\frac{\hat{\gamma}\chi}{\nu}\right)-\frac{\hat{\kappa}\kappa}{\nu}-\frac{\hat{\chi}}{2}\left \langle (s^\star)^2\right\rangle_t+\left \langle s^\star t\right \rangle _t\sqrt{\hat{\gamma}-\left(1+\lambda\right)\hat{\kappa}^2}+\hat{\kappa}\epsilon_0\left \langle s^\star \right \rangle_t \\ \nonumber
&+&\frac{1}{\nu}\left \langle u\left(x^\star \right)\right \rangle_{t,x_0}-\frac{1}{2\nu\chi}\left \langle \left(x-x_0+ t \sqrt{\nu\Omega}+\frac{\kappa}{1+\lambda}\right)^2\right \rangle_{t,x_0} \ .
\end{eqnarray} 
Note that the equation specifying $x^\star$ --- which will prove to be useful shortly --- reads
\begin{equation} \label{eq:uprime}
\chi u^\prime (x^\star)=x^\star - x_0+t\sqrt{\nu\Omega}+\frac{\kappa}{1+\lambda} \ .
\end{equation}

Posing $p=-\hat{\kappa}$ and $\sigma=\sqrt{\hat{\gamma}-\left(1+\lambda\right)\hat{\kappa}^2}$, Eq.~\eqref{eq:h_app} can be rewritten as the following expression:
\begin{eqnarray*}
h\left(\Omega,\kappa,p,\sigma,\chi,\hat{\chi}\right) &=& \frac{\Omega\hat{\chi}}{2}-\frac{\sigma^2\chi}{2\nu}-\frac{\chi\left(1+\lambda\right)p^2}{2\nu}+\frac{p\kappa}{\nu} +\left<\max_{s\ge 0}\left[-\frac{\hat{\chi}}{2} s^2+\left(t\sigma-p\epsilon_0\right) s\right]\right>_t  \\ \nonumber
&+& \frac{1}{\nu}\left \langle \max_{s\ge0}\left[u\left(x\right)-\frac{1}{2\chi}\left(x-x_0+ t\sqrt{\nu\Omega}+\frac{\kappa}{1+\lambda}\right)^2\right]\right \rangle_{t,x_0} \ ,
\end{eqnarray*}
and the saddle point condition (see Eq.~\eqref{eq:sp}) provides us with a set of equations for the order parameters that fully specify the solution of our optimization problem:
\begin{eqnarray} \label{eq:sp}
\Omega &=& \left \langle \left(s^\star\right)^2\right \rangle_t \\ \nonumber
\hat{\gamma} &=&  \left \langle \left [ u^\prime (x^\star ) \right ]^2\right \rangle_{t,x_0} \\ \nonumber
p &=& -\hat{\kappa} = \frac{1}{1+\lambda}\left \langle u^\prime \left(x^\star \right)\right \rangle_{t,x_0} \\ \nonumber
\kappa &=& (1+\lambda)p\chi+\nu\epsilon_0\left \langle s^\star\right\rangle_t \\ \nonumber
\hat{\chi} &=& \sqrt{\frac{1}{\nu\Omega}}\left \langle u^\prime \left(x^\star \right)t\right \rangle_{t,x_0} \\ \nonumber
\chi &=& \frac{\nu}{\sigma}\left \langle ts^\star \right \rangle_t \\ \nonumber
\sigma &=& \sqrt{\hat{\gamma}-\left(1+\lambda\right)\hat{\kappa}^2}=\sqrt{\left \langle \left [ u^\prime (x^\star ) \right ]^2\right \rangle_{t,x_0} -\frac{1}{1+\lambda}\left \langle u^\prime \left(x^\prime\right)\right\rangle_{t,x_0}^2} \ .
\end{eqnarray}

\section{Statistical properties at equilibrium} \label{app:stat_prop}

In this section we derive the statistical properties of the optimal scales of production $s^\star$ and consumption $x^\star$ at equilibrium. Starting from the former, we see from the third term in Eq.~\eqref{eq:heq} that it must be the solution of the following equation
\begin{equation*}
\frac{\partial}{\partial s}\left (-\frac{\hat{\chi}}{2}s^2+st\sigma-p\epsilon_0s\right )_{s = s^\star}=0 \ ,
\end{equation*}
which implies
\begin{equation*}
s^\star=\frac{t\sigma-\epsilon_0p}{\hat{\chi}}.
\end{equation*}
Recalling that $t$ is a standard Gaussian random variable, so must be $s^\star$. Its PDF reads
\begin{eqnarray*}
P(s^\star) &=& \int_{-\infty}^{\infty}\frac{\de t}{\sqrt{2\pi}} \mathrm{e}^{-t^2/2} \ \delta\left(s^\star - \frac{t\sigma-\epsilon_0p}{\hat{\chi}} \right) =
\int_{-\infty}^{\infty}\frac{\de t}{\sqrt{2\pi}} \mathrm{e}^{-t^2/2}\left(\frac{\hat{\chi}}{\sigma}\right)\delta\left(t - \frac{s^\star\hat{\chi}+\epsilon_0p}{\sigma}\right) \\ \nonumber
&=& \left [ 1-\phi\left(p,\sigma\right)\right] \delta(s^\star)+\frac{\hat{\chi}}{\sigma\sqrt{2\pi}}\Theta(s^\star)\exp\left[-\frac{\left(s^\star\hat{\chi}+\epsilon_0p\right)^2}{2\sigma^2}\right] \ ,
\end{eqnarray*}
where $\Theta(\cdot)$ denotes the Heaviside function and
\begin{equation*}
\phi\left(p,\sigma\right)=\frac{1}{2}\operatorname{erfc}\left(\frac{\epsilon_0p}{\sigma\sqrt{2}}\right) \ .
\end{equation*}
From the above results, we can obtain an explicit expression for the average optimal scale of production:
\begin{equation} \label{eq:avg_sstar}
\langle s^\star \rangle = \frac{\sigma}{\hat{\chi}\sqrt{2\pi}}\exp\left[- \left ( \frac{\epsilon_0 p}{\sigma \sqrt{2}} \right )^2 \right]-\frac{\epsilon_0 p}{\hat{\chi}}\phi(p,\sigma) \ .
\end{equation}

Combining Eqs.~\eqref{eq:heq} and~\eqref{eq:uprime}, and following the same line of reasoning of the above derivation, we get the following expression for the PDF of optimal consumption values at equilibrium
\begin{equation*}
P(x^\star|x_0)=\frac{1-\chi u^{\prime\prime}(x^\star)}{\sqrt{2\pi \nu\Omega}}\exp\left[- \frac{1}{2\nu\Omega} \left(x^\star-x_0-\chi u^\prime(x^\star)+\frac{\kappa}{1+\lambda}\right)^2 \right] \ ,
\end{equation*}
which can be integrated against the PDF of initial endowments (Eq.~\eqref{eq:f0_pdf} in our case) to obtain an unconditional PDF $P(x^\star)$. Contrary to the case of scales of productions, with our choice of utility function ($u(x) = \log(x)$) the above expression does not allow to derive an explicit expression for the average $\langle x^\star \rangle$, which has instead to be computed via numerical integration.

The above expressions can be used to obtain the statistical properties of the economy's GDP $G$, which is GET models such as the one used in this work is usually defined as the total market value of all goods in the economy, at equilibrium i.e., $\sum_{c=1}^C p^c [ (x^\star)^c - x_0^c ]$ (price of good $c$ times its net consumption at equilibrium, summed over all goods), normalized to the average price level, i.e.,
\begin{equation*}
G = \frac{\sum_{c=1}^C p^c [ (x^\star)^c - x_0^c ]}{\sum_{c=1}^C p^c} \ .
\end{equation*}
Notably, it can be shown~\cite{de2007typical} that equilibrium prices can be computed as $p^c = u^\prime(x^\star)$.

\bibliography{waste_stat_mech_biblio}

\end{document}